 \documentclass[pdflatex,sn-nature]{sn-jnl}

\usepackage{graphicx}%
\usepackage{multirow}%
\usepackage{amsmath,amssymb,amsfonts}%
\usepackage{amsthm}%
\usepackage{mathrsfs}%
\usepackage[title]{appendix}%
\usepackage{xcolor}%
\usepackage{textcomp}%
\usepackage{manyfoot}%
\usepackage{booktabs}%
\usepackage{algorithm}%
\usepackage{algorithmicx}%
\usepackage{algpseudocode}%
\usepackage{listings}
\usepackage{times}

\begin{document}

\title[The Promise of Gravitational-Wave Astronomy]{The Promise of Gravitational-Wave Astronomy}

\author{\fnm{Parameswaran} \sur{Ajith} \email{ajith@icts.res.in}}

\affil{International Centre for Theoretical Science, Tata Institute of Fundamental Research, Bangalore 560089, India}

\abstract{We review the promise of upcoming gravitational-wave (GW) observations in addressing some of the important open questions in astrophysics and cosmology. A decade after the first direct detection of GWs, the field is entering a transformative phase driven by progressive detector upgrades and the anticipated advent of space-based and next-generation ground-based observatories. These instruments will not only push the distance horizon of GW observations to cosmological redshifts, but will dramatically extend the accessible frequency band, opening entirely new source classes  that lie beyond the reach of current detectors. The source landscape is rich enough that surprises are likely. This review surveys our current understanding across this landscape and offers some perspective on the discoveries that upcoming GW observations are poised to deliver.}


\maketitle

\section{Introduction}\label{sec1}

Gravitational-wave (GW) astronomy has already produced its share of surprises. {The first binary black hole (BBH) event, GW150914~\cite{LIGOScientific:2016aoc}, revealed unexpectedly heavy BHs, while the first binary neutron star (BNS) event, GW170817~\cite{LIGOScientific:2017vwq}, was an extraordinarily rich multimessenger event. Both events were also unusually loud relative to simple statistical expectations.} The electromagnetic (EM) counterparts of GW170817 were observed across the spectrum~\cite{LIGOScientific:2017ync}---from radio to gamma rays---making it one of the most intensively studied objects in modern astronomy. When LIGO~\cite{LIGOScientific:2007fwp,LIGOScientific:2014pky} and Virgo~\cite{VIRGO:2012dcp,VIRGO:2014yos} detectors were originally planned, the most anticipated sources were the inspiral of BNSs~(see, e.g.,~\cite{LIGOScientific:2010nhs}). However, the sample of detected sources is dominated by BBH mergers~\cite{LIGOScientific:2025slb}~\footnote{Indeed, the intrinsic merger rate of BNSs is larger than that of BBHs~\cite{LIGOScientific:2026ctl}. However, since GW detectors are able to detect BBHs out to much larger distances, the observed sample is dominated by them. This point was also well-articulated by some early studies; see, e.g.,~\cite{Flanagan:1997sx}.}. GW observations also uncovered a population of heavy black holes (BHs) with masses $\sim 30 - 100 M_\odot$ that came a surprise to many astronomers~\cite{LIGOScientific:2025slb}. Equally surprising is the observation of compact objects in the putative ``mass gap'' ($\sim 2.5 - 5~M_\odot$) between BHs and neutron stars (NSs)~\cite{LIGOScientific:2024elc,LIGOScientific:2020zkf}. These could be either the lightest BHs or the heaviest NSs ever observed, if not something more exotic. There was also a prior expectation that most of the astrophysical BHs would be highly spinning, due to their accretion history (see, e.g.,~\cite{McClintock:2011zq}). However, LIGO-Virgo observations suggest a large fraction of  BBHs to have modest, or no spins~\cite{LIGOScientific:2025slb}.

The LIGO-Virgo-KAGRA (LVK) Collaboration has announced the detection of over 300 GW signals~\cite{LIGOScientific:2026sit}. Independent analysis of the LVK data has produced over two {dozen} additional detection candidates~\cite{Nitz:2021zwj,Olsen:2022pin,Wadekar:2023gea}. In the third part of the fourth LVK observing run (O4c), 68 significant, non-retracted GW candidates were identified by the LVK low-latency searches \citep{LVK_GraceDB_O4}. A six-month-long intermediate observing (IR1) run is planned to start by the end of 2026~\cite{IR1}. The fifth observing run (O5), which will include LIGO, Virgo and KAGRA~\cite{KAGRA:2020tym} detectors with improved sensitivities, will start in a few years~\cite{abbott2020prospects}, potentially observing thousands of mergers~\cite{Chen:2024gdn,Maity:2025apt}. LIGO-India~\cite{Iyer:2011LIGO-India,Saleem:2021iwi} is expected to join the network in the early 2030s. Beyond the fifth observing run, proposed upgrades to LIGO include $A^{\sharp}$~\cite{HLA-psd-O6} and Voyager~\cite{Adhikari:2019zpy}. The next-generation detectors like Cosmic Explorer~\cite{Reitze_2019, evans2021horizonstudycosmicexplorer} and the Einstein Telescope~\cite{maggiore2020science, hild2011sensitivity} are expected to significantly increase the observing horizon in the high-frequency band ($\sim 10-10^3$ Hz) of GW astronomy, reaching redshifts of $z \sim 10-100$. These detectors may detect millions of binary mergers in coming decades. The proposed high-frequency ($\sim 2 - 4$ kilo Hz) detector NEMO~\cite{Ackley:2020atn} aims to detect the post-merger oscillations of NSs.   

In parallel to these advances in high-frequency GW astronomy, several pulsar timing arrays (PTAs) have announced modest evidence for very low-frequency ($\sim$ nano Hz) GWs~\cite{NANOGrav:2023gor,EPTA:2023fyk,Reardon:2023gzh,Xu:2023wog}. The most promising signal in this band is the stochastic background produced by the inspiral of a large number of supermassive BHs (SMBHs). A joint analysis of the data from different PTAs is currently ongoing, which will hopefully confirm the nature of this signal excess. PTAs are expected to inaugurate nano Hz GW astronomy in the next decade. 

The space-based Laser Interferometer Space Antenna (LISA), with an expected launch in 2035, will open the milli-Hz GW astronomy, observing the merger of SMBHs~\cite{2017arXiv170200786A}. LISA is also expected to observe several extreme mass ratio inspirals (EMRIs), in which NSs or stellar-mass BHs inspiral into SMBHs. LISA is also expected to observe $\sim 10^4$ resolvable white-dwarf (WD) binaries in our galaxy (a small fraction of them are already known through EM observations) and possibly the early inspiral of $\sim 10^2$ stellar-mass binaries involving NSs and BHs. A stochastic background of millions of galactic WD binaries will act as a confusion background for LISA~\cite{LISA:2022yao}. The proposed TianQin~\cite{TianQin:2015yph} detector aims to cover a wider frequency band from milli Hz to deci Hz. 

There are also several concepts for deci-Hz detectors, including space- or Moon-based interferometric detectors~\cite{Kawamura:2020pcg,Kuns:2019upi,Jani:2025uaz,Sharma:2026tff} as well as lunar detectors aimed to measure the mechanical response of the Moon to GWs~\cite{LGWA:2020mma}. These detectors will be able to observe the long inspiral of stellar-mass binaries precisely, providing early warning of their mergers. This will enable astronomers to point their telescopes to the source to catch any EM precursors or prompt emission from the merger. Other potential deci-Hz sources include the merger of binary WDs and intermediate-mass BHs (IMBHs)~\citep{Ajith:2024mie}. The most ambitious deci-Hz detector concepts aim to detect even the primordial GW background~\cite{Kawamura:2020pcg}. 

An especially powerful prospect is \emph{multiband} GW astronomy: the same stellar-mass binary may first be observed during its long inspiral by LISA or a deci-Hz detector and later during merger by a ground-based detector~\cite{Sesana:2016ljz}. Combining observations separated by years and by several decades in frequency can improve measurements of source parameters and merger time and provide early warning for EM observations. Additionally, observations across the nano-Hz, milli-Hz, deci-Hz and ground-based bands will connect different mass scales and evolutionary stages of SMBH binaries, providing a more complete view of their cosmic assembly.

Upcoming GW observations have the potential to transform our understanding across multiple frontiers. They will uncover the population properties of WDs, NSs, and BHs, illuminating the formation and evolution of compact binaries and the physics of stellar evolution and collapse. They will deliver independent measurements of cosmological parameters, contributing to resolving the Hubble tension and to a better understanding of dark energy. They will provide new constraints on the nuclear equation of state (EoS) and potential hints on the nature of dark matter. New signal classes---continuous waves from spinning NSs, bursts from core-collapse supernovae, gravitationally lensed GWs, and a rich milli-Hz landscape of binary WDs, massive BH mergers, and EMRIs---will each open distinct scientific windows, many amplified by multimessenger follow-up. These observations will also enable precision tests of general relativity (GR) and probe the nature of BHs. And there remains the prospect of genuine surprises: cosmic strings, exotic compact objects, ultralight boson clouds around BHs, and wholly unanticipated signals.

This review aims to provide a broad-brush overview of the astrophysical and cosmological sources of GWs, and the major scientific questions we anticipate to address through their observation. {It is useful to keep in mind a loose hierarchy of possible outcomes. Some measurements are \emph{expected}; e.g., precise population measurements of BHs and NSs in binaries. Others address questions whose answers remain genuinely \emph{uncertain}, such as the relative importance of different binary-formation channels. Beyond these lie observations that would be genuinely \emph{surprising}, such as a population of a sub-solar-mass or high-redshift compact objects. Finally, a small set of possibilities would be \emph{revolutionary}: convincing evidence for a non-Kerr compact object, a primordial cosmological background, or new particle physics revealed through GWs. These categories are not sharp---an observation may move from one to another as our theoretical and observational understanding develops---but they provide a useful way of thinking about the discovery space.} We begin with the most well-established sources and their associated science, eventually moving to more speculative but scientifically profound targets. Throughout, we touch upon the surprises that future observations may reveal. The emphasis throughout is on our current understanding and the prospects and challenges for the future, rather than on historical development or comprehensive literature coverage; citations are therefore weighted towards recent articles and reviews that the reader can use as entry points into the primary literature.

\section{Expected sources,  science cases and potential surprises}
\label{sec:sources}

\subsection{Coalescing binaries of black holes and neutron stars}

So far, the only guaranteed sources for ground-based high-frequency detectors are coalescing compact binaries involving BHs or NSs. From the present observations, the estimated merger rate of BNSs in the local universe ($z \sim 0 $) is {$5.1-154.7$ / yr / Gpc$^3$}, while the same for BBHs (NS-BH binaries) is {$27.5-49.4~ (6.7-32.8)$ / yr / Gpc$^3$}~\cite{LIGOScientific:2026ctl}. It is quite likely that the merger rates will evolve with cosmological redshift. In the absence of any high-redshift observations of such mergers, different astrophysical models make different predictions of high-redshift merger rates. 

While there is some uncertainty in the expected detection rates, with the proposed upgrades, current detectors are likely to observe tens of thousands of compact binary mergers (see, e.g.,~\cite{Chen:2024gdn,Maity:2025apt}). This will enable powerful probes of the population properties of BHs and NSs in binaries. Currently, the astrophysical mechanisms that enable the formation of merging compact binaries are largely uncertain. Multiple formation channels are proposed, which include the isolated evolution of massive stellar binaries through a common envelope stage~(see, e.g.,~\cite{Belczynski:2016obo}), chemically homogeneous evolution of isolated binaries~(see, e.g., \cite{deMink:2016vkw}), dynamical formation of binaries in dense stellar environments such as globular or nuclear clusters~(see, e.g.,~\cite{Rodriguez:2015oxa}), binary formation in active-galactic-nucleus disks (see, e.g.,~\cite{Ford:2021kcw}), etc. 

Some of these formation channels make broad predictions on the distributions of source properties. For example, due to the GW radiation reaction, isolated binaries formed at large separations would have circularized by the time they reach the frequency band of ground-based detectors~\cite{Peters:1974}. On the other hand, binaries formed dynamically at close separations could retain some residual orbital eccentricity even when they reach the late stages of their inspiral. Similarly, due to several mass transfer phases during the binary star evolution, compact objects in isolated binaries are likely to have their spins aligned with the orbital angular momentum, while those formed by dynamical interactions need not have any preferred spin orientations. Also, due to the effect of mass segregation, binaries formed through dynamical interactions in clusters are likely to be more massive with near-equal mass ratios. 

One might also see specific imprints of stellar evolution in the population properties of compact objects. For example, massive stars in the main-sequence mass range $\sim 70 - 260 ~M_\odot$ are expected to obliterate themselves through the pair instability supernova (PISN), leaving no remnant.  This should produce a gap in the range $\sim 50-130 ~M_\odot$ in the mass distribution of BHs~\cite{Woosley:2016hmi,Farmer:2019jed}.  

While these basic arguments are well-motivated, in practice it would be difficult to tease out the specific imprints from data. Some of the effects might be too weak to measure --- for example, for a large fraction of dynamically formed binaries, the orbital eccentricity will be reduced to values below the measurable levels by the time they reach the frequency band of ground-based detectors~\cite{Rodriguez:2017pec}. Since the effect of spins is subdominant on the waveform, it is difficult to measure the spin orientations precisely from most signals. {The PISN gap need not remain empty: hierarchical mergers can produce second-generation BHs whose masses lie in the nominal gap, and uncertainties in stellar evolution can shift its boundaries.} Finally, multiple channels might be contributing to the observed samples. 

Despite these challenges, the large number of expected observations combined with detailed modeling of the physical and astrophysical processes involved, should give us powerful insights into the formation of coalescing compact binary systems, potentially resulting in surprises.  Even current GW observations have enabled us to reconstruct the mass distribution of BHs and NSs in coalescing binaries, although with significant statistical uncertainties~\cite{LIGOScientific:2026ctl}. There are already hints of interesting features in the reconstructed mass function, including a peak at $\sim 10 M_\odot$ and a change of slope at $\sim 35 M_\odot$. Equally surprising are the observed ``lower-mass-gap objects'' --- compact objects in the mass range $2-3 M_\odot$, which could be the heaviest NSs or the lightest BHs ever observed. The stochastic GW background produced by a large number of compact binaries merging at different redshifts will be detectable in the next few years~(see, e.g., \cite{LVK:O4a_stoch}). The amplitude and shape of the spectrum will have imprints of the population properties of compact binaries, such as their masses, redshift evolution and orbital eccentricities. By itself the background is unlikely to identify formation channels uniquely, because different astrophysical channels can produce degenerate spectra; rather, it will provide complementary constraints on portions of the population that are individually unresolved~\cite{Mukherjee:2021ags,Bavera:2021wmw,Zhao:2020iew}.  

As the detector sensitivity improves, we will observe a few events with high signal-to-noise ratios (SNRs), which will enable precision tests of GR and the true nature of compact objects. The recent BBH merger GW250114 has already given us a glimpse of such possibilities~\cite{GW250114_testGR,GW250114_testGR_AreaThm}. While GR has been tested well by a variety of laboratory experiments and astronomical observations, GW observations will probe an entirely unexplored regime of extreme velocities and strong gravitational fields~(see, e.g., \cite{Berti:2015itd}). 

Observations of BNSs will enable us to probe the EoS of nuclear matter at extreme densities that cannot be probed by laboratory experiments. The tidal deformation of NSs, which will leave an imprint on the GW signal, contains information of the EoS~\cite{Flanagan:2007ix,Hinderer:2009ca}. Precise measurement of the mass $m$ and tidal deformability $\Lambda$ of several NSs can be used to constrain the EoS, similar to what is attempted from the mass-radius relation of NSs~(see, e.g., \cite{Ozel:2016oaf,Koehn:2024set}). If the merger remnant is a NS, it will be significantly hotter, making temperature an important EoS parameter. The post-merger GW signal will contain imprints of the hot nuclear EoS, along with that of any phase transitions (e.g., to quark matter)~\cite{Bauswein:2019skm}.

Coalescing compact binaries are \emph{standard sirens} --- luminosity distance to the source can be accurately measured from GW signals without relying on any distance ladders~\cite{Schutz:1986gp,Holz:2005df}. However, {for vacuum BBH waveforms the cosmological redshift is degenerate with the total mass} --- a reflection of the absence of any preferred scale in GR. {Additional source-frame scales, such as tidal effects in NSs, can in principle help break this degeneracy~\cite{Messenger:2011gi}.} However, the cleanest way to obtain the redshift is from an EM counterpart, as successfully demonstrated in the observation of the first BNS merger event GW170817~\cite{LIGOScientific:2017ync}. Observation of several such multi-messenger sources will allow us to populate the Hubble diagram, enabling independent measurement of cosmological parameters. Since compact binaries are well-modeled in GR, the systematics associated with GW observations are typically well understood. This will be an important clue towards addressing some of the current tensions in cosmology. Even mergers without EM counterparts could be used to measure cosmological parameters by associating the mergers statistically with host galaxies~\cite{Schutz:1986gp,DelPozzo:2011vcw}. {Well-understood features in the compact-object mass distribution, such as the maximum NS mass~\cite{PhysRevD.48.4738,Taylor:2011fs,Farr:2019twy}, could provide an alternative way to infer redshift. However, using features in the \emph{inferred} mass spectrum of BBHs~\cite{Ezquiaga:2022zkx} would require accurate modeling of the underlying population and its evolution with redshift.}  Modest measurements of cosmological parameters have already become possible~\cite{LIGOScientific:2026uyd}. In the future, precise and accurate measurements of cosmological parameters, including the EoS of dark energy, is anticipated from the next-generation GW detectors~\cite{Branchesi:2023mws}.

A small fraction ($\sim 0.1 - 0.5\%$) of merger events will undergo strong gravitational lensing by intervening galaxies and clusters, producing multiple images that arrive at the detector with time delays of minutes to months. The first detection of lensed GWs is expected in the next few years as ground-based detectors observe thousands of GW signals~\cite{Barsode:2025agk,Maity:2025apt}. Future observations of lensed GWs could enable precise localization of binary BH mergers~\cite{Hannuksela:2020xor}, provide new probes of cosmology and dark matter~\cite{Jana:2022shb,Jana:2024dhc}, and enable new tests of GR~\cite{Goyal2021,2022arXiv221101272M}. In addition, compact lenses with gravitational radius comparable to the wavelength of GWs will produce wave optics effects on the GW signal~\cite{2003ApJ...595.1039T}---something that is practically impossible to observe in the lensing of EM waves. {The event rates for wave optics lensing are likely to be low unless a significant fraction of dark matter is in the form of such objects. The non-observation of lensed GWs~\footnote{Although the massive BBH event GW231123 appears to be an outlier in recent microlensing searches~\citep{Hu:2025lhv, Goyal:2025eqo, Shan:2025dcd, Chakraborty:2025pxt,Cheung:2026pky}, its lensing interpretation remains unconfirmed.} has already produced some modest constraints on the fraction of dark matter in the form of massive compact objects, such as primordial BHs in the mass range $\sim 10^2-10^8 M_\odot$ \cite{Urrutia:2021qak,Basak:2021ten,LIGOScientific:2023bwz,Barsode:2024wda}.} Upcoming GW observations will either tighten these constraints, or enable a groundbreaking positive detection. 

GW observations could also potentially probe particle dark matter. One idea is that NSs could capture dark matter particles. NSs harboring dark matter cores would appear as outliers in the $m-\Lambda$ curve~\cite{Ellis:2018bkr}. The basic idea is very compelling---while terrestrial dark matter detectors have masses of a few tons and operate only for decades, NSs are $\sim 10^{27}$ times more massive and $\sim 10^{8}$ times longer lived. If the accumulated mass exceeds a critical threshold, the NS core undergoes gravitational collapse, which gets transmuted into a BH of roughly the original NS's mass~\cite{PhysRevD.40.3221,Dasgupta:2020mqg}. The non-observation of such low-mass BHs ($0.5-2~M_\odot$), has been used to constrain some dark matter models~\cite{Bhattacharya:2023stq}; a positive detection in the future will hint at the existence of primordial or transmuted BHs---both equally exciting. 

Another possibility of probing dark matter is through the environmental effects on compact binary inspirals~\cite{Bertone:2024rxe}. If the astrophysical environment of the binary is sufficiently dense (e.g., due to dark matter overdensity), the loss of the orbital energy due to dynamical friction would make the binary inspiral faster, as compared to the same in a vacuum environment~\cite{Eda:2014kra}. This effect could be potentially detected in binaries with large mass ratios~\cite{Kavanagh:2020cfn}.

Binaries involving NSs are potential multimessenger sources, as illustrated spectacularly by GW170817~\cite{LIGOScientific:2017vwq,LIGOScientific:2017ync}. That event established the connection between BNS mergers and (at least a fraction of) short gamma-ray bursts (GRBs), {provided compelling evidence that BNS mergers are sites of r-process nucleosynthesis}, enabled an independent measurement of the Hubble constant, provided independent constraints on the nuclear EoS, and placed exquisite bounds on the speed of GWs.  Upcoming EM telescopes will be able to detect kilonovae out to $z \sim 0.3-0.4$, while GRB afterglows could be detected out to $z > 1$, for viewing angles $\lesssim 15^o$~\cite{Branchesi:2023mws,Bisero:2025tkw}. Current forecasts suggest that only a small fraction of BNS events will have observable EM counterparts, owing to the interplay of sky localization, kilonova luminosity, and GRB viewing angle. Thus, $\sim 10^2$ golden events per year will drive the multimessenger science in the era of next-generation GW detectors. Even these would enable an incredible amount of multimessenger science. Note that these forecasts are made with limited observed samples; there are many ways for the universe to hide its surprises. 

\subsection{Spinning neutron stars}
\label{sec:spinnningNS}

Out of the $\sim 4000$ currently observed pulsars, $\sim 30\%$ are spinning rapidly enough to produce continuous GWs in the frequency band of ground-based detectors~\cite{Manchester2005ATNF}. These are nearly monochromatic waves with a very small frequency derivative, due to the slow spin down of the NS due to the emission of EM or gravitational radiation. These signals get Doppler modulated due to the orbital and spin motion of the Earth, as well as any motion of the NS. These signals are much weaker than those from compact binary mergers; however, these persistent signals could be integrated over months to years to accumulate SNR. Our galaxy has an estimated $\sim 10^8-10^9$ NSs. However, the distribution of the equatorial ellipticity $\epsilon$ of NSs, which determines the strength of the GWs, is poorly understood. {Current searches place source-dependent upper limits on the GW amplitude and ellipticity; for the most favorable known pulsars these limits reach well below $10^{-8}$, while the constraints are substantially weaker for many other targets~\cite{LIGOScientific:2025kei}.} Rapidly spinning NSs with $\epsilon \sim 10^{-8} - 10^{-9} $ will produce continuous GWs observable by the upgraded present-generation detectors or next-generation detectors~\cite{Riles:2022wwz,Owen:2025ecd}.  Non-detection in the next decade would begin to raise serious doubts about current theories of the formation of millisecond pulsars~\cite{Owen:2025ecd}. The planned deci-Hz detectors could detect some of the slowly spinning NSs, such as magnetars~\cite{Pagliaro:2025qrr}. 

Observation of continuous GWs from NSs could yield insights into the GW generation mechanisms~\cite{Owen:2025ata}, the NS structure and EoS, although we have to overcome a number of practical difficulties (see, e.g.,~\cite{Jones:2024npg}). Such observations could also enable interesting tests of GR by measuring, or constraining, the presence of GW polarizations that are absent in GR (scalar or vector modes), but might be present in alternative theories of gravity~\cite{Will:2014kxa,Isi:2017equ}. More speculatively, there is a small chance to discover a spinning NS behind the galactic center that could be gravitationally lensed by the supermassive BH~\cite{Basak:2022fig}. If detected, this would enable powerful probes of the galactic center~\cite{Basak:2022fig,PhysRevD.109.024064}.

\subsection{Core-collapse supernovae}

Collapse of massive stars and the subsequent oscillations in the newly formed NS will produce a burst of GWs. However, recent 3D simulations suggest the core collapse to be nearly spherical, and most of the energy is lost into neutrinos. {The resulting GW signal is expected to be weak and highly dependent on the explosion dynamics and progenitor rotation. Galactic events would be prime targets, while next-generation detectors could extend sensitivity to more distant events for sufficiently favorable and energetic emission scenarios~\cite{Muller:2026ofz}.} The estimated supernova rate in our galaxy is $\sim 1-3$ per century---an event is possible but far from guaranteed on decadal timescales.

Even though core-collapse supernovae are rare, they are exceptional multi-messenger events. Complementary to the photons that are produced in the outer edge of the exploding star, GWs carry information about the dynamics of the central engine, while neutrinos carry information about thermodynamic conditions. Such multi-messenger observations will probe the central engine in a completely new way, extracting information about the supernova explosion mechanism as well as the rotation and the structure of the innermost regions of the star (see, e.g.~\cite{Muller:2026ofz}).

\subsection{Binary white dwarfs}

Our galaxy has an estimated $\sim 10^{8}$ WD binaries producing nearly continuous GWs in the frequency range $10^{-4} - 10^{-3}$ Hz. The incoherent superposition of these signals will produce an unresolvable ``confusion'' background in LISA. However,  $\sim 10^4$ of them are in high-enough frequencies (a few milli Hz) where their inspiral will be measurable by LISA, rendering them individually resolvable. LISA will be able to identify their sky locations accurately, which will allow for follow-up EM observations. Some of them, called \emph{verification binaries}, are already well-characterized through EM observations, with known positions, periods, and sometimes masses~\cite{LISA:2022yao}. 

Multi-messenger observations could address key scientific questions related to formation and evolution of binary WDs, such as the stability and efficiency of mass transfer, common envelopes, tides and stellar angular momentum transport, etc. The observed number of WD binaries could be used to estimate the WD merger rate in the galaxy, which can be compared with the rate of Type-Ia supernova to test the hypothesis that they are produced by WD mergers~\cite{LISA:2022yao,Ajith:2024mie}. 

Deci-Hz GW detectors will also observe mergers of WD binaries. The galactic merger rates are likely to be too low ($\lesssim 10^{-2}$ / yr) to result in a detection during the observation time of $\sim 10$ yrs.  However, the proposed deci-Hz detectors should be able to observe WD mergers out to $\sim 100$ Mpc, potentially detecting dozens of WD mergers.  If a WD merger is confidently associated with a Type-Ia supernova (which might be difficult due to the uncertain and possibly long time delay between them), this opens up an interesting possibility of calibrating the supernova distance, as the luminosity distance could be measured directly from the GW signal~\cite{Ajith:2024mie}. Such a direct calibration of Type-Ia supernovae, independent of the cosmic distance ladder, will significantly reduce the systematics associated with estimation of cosmological parameters using supernovae, and will have direct bearing on resolving the Hubble tension~\cite{DiValentino:2021izs}. 

\subsection{Binaries involving massive black holes}

In hierarchical structure formation, SMBHs are expected to grow from lower-mass ``seed'' BHs through gas accretion and successive mergers. But the masses, formation sites and abundance of the initial seeds remain poorly known~(see, e.g.,~\cite{Volonteri:2021_smbh_rev}). Broadly, seed scenarios range from light remnants of the first generations of stars to much heavier seeds produced by direct collapse or other mechanisms. Their subsequent growth depends on poorly constrained accretion histories, galaxy mergers and the efficiency with which BH binaries form and coalesce.

LISA is expected to observe mergers of BHs in the mass range $\sim 10^4-10^7 M_\odot$ to very high redshift, potentially out to $z\sim20$~\cite{LISA:2022yao}. The masses, redshifts, spins and mass ratios of these mergers will directly probe the assembly history of massive BHs. Different seed and growth scenarios predict different merger populations, especially at high redshift, where EM observations are most challenging. Thus, the absence or presence of particular populations may be as informative as the total detection rate. A substantial population of IMBHs in the range $\sim10^2-10^5 M_\odot$ would itself be an important discovery, filling the poorly explored gap between stellar-mass and supermassive BHs. Deci-Hz detectors can access IMBH mergers in the mass range  $\sim 10^2 - 10^4 M_\odot$~\cite{Ajith:2024mie}, while next-generation ground-based detectors probe still lower masses ($\sim 10-10^3 M_\odot$) out to high redshifts~\cite{Branchesi:2023mws,Evans:2021gyd}.

At still larger masses, PTAs are sensitive to the collective signal from binaries containing $\sim10^7-10^9 M_\odot$ BHs. The amplitude and spectral shape of this nano-Hz background, and eventually individually resolved binaries, will constrain the merger history and environments of SMBH binaries. Taken together, PTA, LISA, deci-Hz and ground-based observations could therefore trace BH assembly across several orders of magnitude in mass and over much of cosmic history, in some cases aided by EM counterparts. This is a particularly promising arena for surprises because present models extrapolate uncertain seed formation, accretion and galaxy evolution into epochs that are only weakly constrained observationally~\cite{LISA:2022yao}. These sources can also be used as standard sirens and as probes of modified GW propagation and dark energy~\cite{LISA:2022kgy}.

Galactic nuclei hosting SMBHs are also expected to contain dense populations of stars and compact objects. A stellar-mass BH or NS captured onto a relativistic orbit can produce an EMRI, accumulating $\sim10^4-10^5$ observable orbital cycles before plunging into the central object. Such signals can map the multipolar structure of the central spacetime with high precision and test whether it is consistent with the Kerr geometry predicted by GR. A statistically convincing inconsistency would be revolutionary. EMRIs may also reveal otherwise inaccessible properties of galactic nuclei---including stellar cusps, gas and possible dark-matter overdensities---through environmental perturbations to the waveform~\cite{LISA:2022yao,LISA:2022kgy}.

\subsection{Cosmological sources}

GW observations can help us to probe the late universe (including the nature of dark matter and dark energy) in multiple ways---through GW standard sirens, through the lensing of GWs, and, more speculatively, by providing an independent calibration for Type Ia supernovae. GW observations will also constrain the abundance of primordial BHs in some mass ranges ($\sim 0.1 - 10^6 M_\odot$), thus indirectly probing the early universe. 

{More fundamentally, GWs can carry information from epochs that are inaccessible to EM observations. Physical processes in the early universe---including inflation, first-order phase transitions and topological defects---can generate stochastic GW backgrounds~(see, e.g.,~\cite{Caprini:2018mtu}). Once produced, these waves propagate essentially freely, potentially preserving a fossil record of the expansion history and high-energy particle physics of the early universe. The principal observational challenge is to separate such a cosmological signal from astrophysical foregrounds generated by compact binaries.}

{The background generated by standard slow-roll inflation is expected to be very weak at frequencies accessible to currently planned interferometers. Ambitious deci-Hz concepts have nevertheless been designed with the goal of measuring and subtracting compact-binary foregrounds sufficiently accurately to observe the cosmological  signal~\cite{Corbin:2005ny,Kawamura:2020pcg}. For the inflationary background, the most promising near-term route remains its imprint on the polarization of the cosmic microwave background~\cite{Kamionkowski:CMBRev_2016}. A direct GW detection, however, would provide a qualitatively new probe of the inflationary epoch.}

{First-order phase transitions provide another especially compelling target. Such transitions can occur when symmetries are broken in extensions of the Standard Model, producing GWs through expanding bubbles, sound waves and turbulence in the primordial plasma. The characteristic frequency today is related to the energy scale of the transition, allowing instruments in different frequency bands to probe particle physics far beyond laboratory energies~\cite{Athron:2023fopt}. A detection would therefore be not merely an improved cosmological measurement but evidence for new high-energy physics.}

{Topological defects provide a third class of targets. Cosmic strings, if formed in an early-universe phase transition, generate a stochastic background---and potentially individual bursts---over an exceptionally broad range of frequencies. Current PTA measurements already constrain parts of the parameter space, while combining PTA, LISA and ground-based observations would test the broadband spectra predicted by different string models~\cite{Sousa:2024cosmicstrings}. A convincing detection of a primordial background from any of these mechanisms would be revolutionary: unlike most sources discussed above, it would give us direct observational access to physical processes in the universe before the formation of stars and galaxies.}

\subsection{Exotica}

Further along the spectrum lie sources and phenomena whose existence is speculative, but whose detection would have profound implications for fundamental physics. {A particularly clean surprise would be the detection of a compact object with a confidently measured mass below the range accessible to ordinary stellar remnants. In standard stellar evolution, BHs are not expected below roughly a solar mass. Consequently, a subsolar-mass BH would point to a non-standard formation mechanism, with primordial BHs providing the most widely discussed possibility~\cite{Bagui:2025pbhreview}. GW searches have already placed limits on subsolar-mass binary mergers, and future detectors will explore much larger volumes. The scientific value of a positive detection would be disproportionate to the number of events: even a small population could provide direct evidence of compact objects formed outside conventional stellar evolution and potentially a link to the nature of dark matter.}

There are also proposals on how GW observations could probe particle physics. One idea is by detecting continuous GWs produced by a cloud of {ultralight bosons} around a spinning BH. Such bosons, e.g., axions and dark photons, are predicted in many extensions to the Standard Model and are potential dark matter candidates. If they exist, with Compton wavelength comparable to the gravitational radius of the BH, they will form a dense boson cloud around the BH through the \emph{superradiance} phenomenon~\cite{Brito:2014wla}. The cloud will extract energy and angular momentum from the BH, causing it to spin down. The resulting GWs are similar to the continuous GWs from spinning NSs~(Sec.~\ref{sec:spinnningNS})~\cite{LIGOScientific:2021rnv}. If such fields exist in the appropriate mass range, these GW signals could be detectable by next-generation instruments, providing direct evidence for physics beyond the Standard Model. 

Another means of probing particle physics is by observing ultracompact objects---{horizonless objects whose compactness can approach that of BHs.} The possible existence of ultracompact objects such as boson stars, fermion stars, gravastars, dark matter stars, etc. have been speculated in the literature~\cite{Giudice:2016zpa}. Since they could have masses and sizes similar to that of BHs, they could mimic BHs in their GW signals as well as EM signatures such as shadows. However, {their multipole moments need not satisfy the Kerr relations, and unlike a classical BH in GR they may possess a non-zero tidal response. Spin-induced multipoles and tidal effects can therefore leave characteristic imprints on the GW signal}~\cite{Krishnendu:2017shb,Johnson-Mcdaniel:2018cdu}. The non-observation of such imprints has put modest constraints on the parameter space of some BH mimickers~\cite{LIGOScientific:2021sio,Krishnendu:2025rud}. A positive detection of such objects would be revolutionary. 

At the outermost edge of speculation lie the possibilities of non-compact extra spatial dimensions, which would modify the amplitude scaling of GWs with distance. The idea is that GWs can propagate through the ``bulk'' (all spatial dimensions), while EM waves are confined to the ``brane'' (the usual three spatial dimensions). This would manifest as an apparent discrepancy in the distance estimate of a multimessenger source using GW and EM signals. {The multimessenger event GW170817 has already placed strong constraints on phenomenological models in which GW amplitudes leak into extra dimensions, as well as on anomalous GW propagation more generally}~\cite{Pardo:2018ipy,LIGOScientific:2018dkp}. 

It would be premature to assign significant probability to any of these; the theoretical frameworks exist, but the underlying physics remains unconfirmed by any other means. They are best understood not as predictions but as illustrations of how broad the discovery space of GW astronomy could be. 

\section{Concluding remarks: The element of surprise in GW astronomy}

The history of observational astronomy is a history of being surprised. Every new EM window---radio, X-ray, gamma-ray---the universe has responded not merely with clearer versions of what was expected, but with entirely new questions and phenomena that challenged and reshaped our understanding. Examples include pulsars, quasars, the accelerating expansion of the universe and fast radio bursts. Many of these were not anticipated with any confidence before their discovery, and each forced profound revisions in astrophysics and sometimes in fundamental physics. GW astronomy, still in its infancy, has already delivered its own share of surprises. The first BBH and BNSs events were unusually loud contrary to prior statistical expectations, and the first BNS event itself inaugurated multimessenger astronomy triggered by GW observations. These, along with the unexpected BH masses, objects in the lower mass gap, and the GRB and kilonova associated with GW170817, marked the spectacular first decade of GW astronomy.  

Whether GW astronomy will prove as surprising as its EM predecessors is, however, an open question. One important difference is that most GW sources are, in a broad sense, simpler than EM sources. GWs are generated by coherent bulk motions of matter and spacetime, and depend primarily on the macroscopic properties of the system. For example, a merging binary is governed by a small number of parameters such as the masses and spins of the compact objects, and GR makes precise predictions for the emitted waveform. EM astronomy, by contrast, encounters the full complexity of plasma physics, nuclear physics, magnetic fields, radiation processes and chemistry, apart from gravity. It is this complexity that often produces unexpected phenomena. The cleaner physics of GW sources is a strength for precision measurement, but it may also narrow the space within which truly unexpected behavior can hide. Surprises, when they come, may be more subtle; e.g., small deviations in waveforms and unexpected population statistics;  rather than the dramatic discoveries of an entirely new class of objects.

That said, there remain ample reasons for optimism. Next-generation detectors in ground and space will not only expand the distance horizon of GW observations, but will dramatically extend their frequency band, accessing source classes entirely beyond current reach. In EM and particle astronomy, new frequency/energy bands have historically uncovered new phenomena. There are two distinct ways in which future GW observations can generate surprises. The first is \emph{new sensitivity}: familiar source classes will be observed to much larger distances and into poorly explored cosmic epochs, where unexpected populations or evolutionary trends may emerge. The higher sensitivity will also enable higher precision: because GR predicts many GW signals with remarkable accuracy, small discrepancies in waveforms can be apparent. In this sense, the simplicity of GW sources is not only a limitation on the space of surprises; it is also what makes subtle surprises identifiable. The second is \emph{new frequency windows}: PTAs, LISA and possible deci-Hz observatories access source classes and evolutionary stages that ground-based detectors simply cannot see. 

The most honest statement one can make now is that we do not know how surprising GW astronomy will ultimately prove to be. The sources reviewed here represent our best current map of the territory, but the map may be incomplete in ways we cannot anticipate.  What we can say with confidence is that the universe has rarely failed to reward a genuinely new way of looking at it. The coming decade of GW observations may not overturn our models as dramatically as the discovery of pulsars or dark energy overturned theirs. But it would be equally imprudent to assume it will not.

\bmhead{Acknowledgments}

I am grateful to many colleagues from the LIGO-Virgo-KAGRA collaboration, the ICTS Astrophysics \& Relativity group, and the wider GW community for numerous discussions that shaped my understanding. I am especially grateful to K. G. Arun, Bala Iyer and Surhud More for their useful comments on the manuscript. My research is supported by the Department of Atomic Energy, Government of India, under project nos. RTI4019 and RTI4013. I also acknowledge support from the ICTP through the Associates Programme and from the Simons Foundation through the grant (Record ID: SFI-MPS-T-Institutes-00012057, AD).

\bibliography{references}
\end{document}